\documentclass[useAMS,usenatbib,usegraphicx]{mn2e}

\title[What drives the radio slopes in radio quiet quasars?]{
What drives the radio slopes in radio quiet quasars?}
\author[A. Laor, R.~D. Baldi \& E. Behar ]
{Ari Laor$^1$\thanks{E-mail: laor@physics.technion.ac.il (AL); 
R.Baldi@soton.ac.uk (RDB);
behar@physics.technion.ac.il (EB) },
Ranieri D. Baldi$^{2}$\footnotemark[1]
and
Ehud Behar$^{1}$\footnotemark[1] \\ 
$^1$Physics Department, Technion, Haifa~32000, Israel \\
$^2$School of Physics and Astronomy, University of Southampton, Southampton, SO17 1BJ, UK}

\begin{document}
\maketitle


\pagerange{\pageref{firstpage}--\pageref{lastpage}} \pubyear{2002}

\maketitle

\label{firstpage}

\begin{abstract}
The origin of the radio emission in Radio Quiet (RQ) quasars  is not established yet.
Important hints can be provided by the spectral slope, and its relation
to other emission properties. We compiled the 
highest resolution 5 and 8.4 GHz {\em VLA} observations 
available of the RQ optically selected PG quasars at $z<0.5$. 
We derive the 5-8.4~GHz spectral slope, $\alpha_R$, for 25 of the 
complete and well studied sample of 71 RQ PG quasars. We find a highly significant correlation 
of $\alpha_R$ with $L/L_{\rm Edd}$,
where high $L/L_{\rm Edd}(>0.3)$ quasars have a steep slope ($\alpha_R<-0.5$),
indicative of an optically thin synchrotron source. In contrast,
lower $L/L_{\rm Edd}(<0.3)$ quasars generally have a flat slope ($\alpha_R>-0.5$),
indicative of a compact optically thick synchrotron source. Flat $\alpha_R$ quasars
also have a low Fe~II/H$\beta$ line ratio, and a flat soft X-ray slope.
The 16 Radio Loud (RL) PGs do not follow the RQ quasar set of correlations,
and their $\alpha_R$ is set by $M_{\rm BH}$, suggesting that the radio
emission mechanisms in RL and RQ quasars are different.
A possible interpretation is that high $L/L_{\rm Edd}$ RQ quasars
produce a strong outflow and an associated optically thin synchrotron emission. 
In lower $L/L_{\rm Edd}$ RQ quasars, the strong outflow is missing, 
and only a compact optically thick radio source remains, possibly
associated with the accretion disk coronal emission. A systematic study of RQ quasars
at higher frequencies, and higher resolution, can test whether a compact flat source indeed 
resides in the cores of all RQ quasars, and allow the exploration of its relation with the coronal X-ray emission.

\end{abstract}


\section{Introduction}

The radio emission in Radio Quiet Quasars (RQQ) may be produced by a variety of mechanisms. 
It could be optically thin synchrotron and free-free emission associated with star formation on the host galaxy 
scale (e.g. \citealt{Padovani11, Condon13, Ke16, Brown17, Taba17}, {\em cf.} \citealt{White15, Zakamska16}), 
synchrotron and free-free emission from an AGN driven outflow (e.g. \citealt{Mundell00, 
Blustin09, Jiang10, Steenbrugge11, Faucher12, Zakamska14, Nims15}), 
optically thick synchrotron emission from a compact source, such as
an accretion disk corona (\citealt{LB08, RL16}, hereafter LB08, RL16), or synchrotron emission on various scales 
from a low power jet
\citep{Falcke95, Wilson95}. 

The radio spectral slope carries diagnostic power on the emission mechanism.
Optically thin free-free has a spectral slope of $\alpha\simeq -0.1$ 
($=d\log f_{\nu}/d\log \nu$). Optically thin synchrotron emission 
is characterized by  $\alpha< -0.5$. Optically thick free-free emission becomes a blackbody, where
$\alpha=2$. Optically thick synchrotron has $\alpha> -0.5$, and
can increase up to $\alpha=2.5$ (e.g. \citealt{RL79}, RL16). 
Free-free absorbed synchrotron emission
can produce $\alpha>2.5$ \citep{Levinson95,BL98}. Below we define a spectrum as
flat when $\alpha> -0.5$, and steep when $\alpha< -0.5$. 

A handful of {\em VLA} studies of the radio spectral slope of RQQ are available
\citep{AB88,Barvainis96,Kukula98,Barvainis05}. These studies
indicate that close to a half of RQQ show a flat spectrum
at 5 to 8.4~GHz, suggesting a compact optically thick synchrotron source
(a fraction of 40\% is found by \citealt{Barvainis96}, using mostly C and D arrays; 
7/16 in \citealt{Kukula98}, using the A-array; and
5/11 in \citealt{Barvainis05}, using various arrays).
 Optical depth estimates (LB08),
and Radiative transfer calculations (RL16), imply that an optically thick source at the given
frequencies and luminosities, is 
smaller than $\sim 0.1-1$~pc.
A handful of nearby X-ray bright RQ AGN with a steep $\alpha$, 
generally show a spectral flattening towards the mm regime (95~GHz, \citealt{Behar15, Behar18}).
The flattening at high frequencies indicates that a compact optically thick radio 
source resides in objects which are dominated at lower frequencies 
by extended optically thin emission.

A compact radio source on pc scale, or smaller, is also indicated by the handful of {\em VLBI}
observations of RQQ \citep{Blundell98, Caccianiga01, Ulvestad05a, Doi13, Panessa13},
which yield lower limits on the brightness temperature of $T_B>10^8-10^9$~K, 
which excludes thermal radio emission. 
A compact source is also indicated by the observed variability sometimes observed 
in a handful of sources (\citealt{Wrobel00, Anderson05, Barvainis05, Mundell09, Jones11, Baldi15},
cf. \citealt{Jones17}).

A compact flat spectrum optically thick synchrotron source, observed in about
half of the RQQ, may in fact be present in all RQQ, if it is an inevitable part 
of an accretion disk magnetised coronal 
emission \citep{Field93, DCF97, MMF00}. The coronal emission may extend to the
radio regime if the magnetised plasma extends to a fraction of a pc, e.g. due to
coronal mass ejections (LB08), or due to spatially extended disc corona.
Such a compact component produces a flat spectral slope (e.g. RL16), but
it apparently dominates the observed radio emission only in about half of RQQ. Why is 
the radio emission in the other half of RQQ dominated by a more extended optically 
thin synchrotron source?

Type 1 AGN show a large variety of correlated emission line properties, most notably the
so-called eigenvector 1 (EV1) set of correlations (\citealt{BG92} and citations thereafter).
High EV1 objects generally have relatively narrow broad lines,
specifically narrow H$\beta$,  
strong Fe~II emission, weak [O III] emission,
and asymmetric line profiles with excess blue wing emission. 
These correlated properties extend to the UV emission lines, 
with enhanced N~V, Si~IV+O~IV], Si~III], C~II, and O~I line emission, and 
suppressed C~IV emission \citep{Wills99}. The set of correlations extend to the X-ray
regime, where a narrower H$\beta$ is associated with a
steeper soft X-ray slope \citep{Laor94, Boller96, Laor97, Grupe99, Grupe04}, and also
a steeper hard X-ray slope \citep{Brandt97, Reeves00, Porquet04, Piconcelli05, Shemmer06}. 
A higher EV1 is likely associated with a higher luminosity in Eddington units, $L/L_{\rm Edd}$ \citep{BG92, Sulentic00,
Boroson02}. The physical mechanism which may drive some of the above correlations, 
is enhanced gas metallicity 
\citep{Shemmer02, Shemmer04}, and enhanced outflows \citep{Shin17} in 
higher $L/L_{\rm Edd}$ objects.

Is the radio spectral slope in RQQ also part of the EV1 set of correlated properties? 
If yes, it may provide important hints on their radio emission mechanism.
Also, do the radio slopes of RLQ and RQQ show a similar set of correlations? 
as may be expected if RQQ are scaled down versions of RLQ.  
To answer these questions, we explore the correlations
of the radio spectral slopes and other emission properties of a well studied sample of quasars,
and compare the results for identically selected RQQ and RLQ.

Below we find that the radio slope in RQQ is indeed part of the EV1 set of correlations, 
where narrow H$\beta$, or equivalently high $L/L_{\rm Edd}$ objects, 
have a steeper $\alpha$. In contrast, RLQ
do not follow these correlations. Instead, their $\alpha$
gets steeper with increasing black hole mass, $M_{\rm BH}$. This contrast suggests that RQQ are not just scaled
down RLQ. Their radio emission is driven by a different physical mechanism, possibly
an outflow which is more prevalent in high $L/L_{\rm Edd}$ objects.

The paper is organised as follows.
In section \S 2 we describe the sample and the observations used, in section \S 3 we
present the results of the correlation analysis, in \S 4 we discuss the results and
in \S 5 provide the main conclusions.

\section{The sample}

Our purpose is to derive the spectral slope of the quasar radio emission, excluding 
as much as possible contamination from 
the host galaxy radio emission, which may be significant in RQQ. 
We therefore exclude single dish observations, 
and use only the highest resolution available {\em VLA} observations,
preferably with the A configuration. To maximise the angular resolution,
we use the higher frequencies which are commonly available, specifically
the C and X bands at 5 and 8.4 GHz. These criteria provide angular 
resolutions of 0.4 and 0.24 arcsec respectively. We further select low redshift objects,
with a typical $z\sim 0.1$ (see below), which corresponds to physical 
integration radii of 370~pc and 220~pc respectively.
These radii are generally well below the half-light radii of galaxies
(e.g. \citealt{Faber97}), and we therefore expect that most of the  
host galaxy radio emission related to stellar processes, is excluded. 
The observed radial extent of the radio
emission in RQQ is indeed generally well below the kpc scale (e.g. \citealt{Ke94}),
suggesting that host contamination is not a major issue. However, star formation contribution 
to the radio emission on the unresolved sub-kpc core scale may still be present.
In some RQQ significant radio emission is present on larger scales, which may be
host related, but may also be AGN related. In this study we focus on the unresolved
core emission, i.e. below a few hundred pc, and ignore the larger scale 
resolved emission. 

We use the PG quasar sample \citep{Schmidt83}, comprised of the brightest 114
quasar in one quarter of the sky, residing above the Galactic plane, and generally 
above the celestial equator.
This sample, although defined 35 years ago, is still the only well define and
complete sample of bright quasars, which satisfies two properties which are 
critical for this study. 
First, the selection
criteria (bright, point-like, blue objects) are independent of the radio properties,
thus the radio properties are not directly biased by selection effects. 
Second, most of the sample ($\sim 90$\%) is detected in the radio at 5~GHz 
\citep{Ke89}.
The sample is therefore well defined, unbiased, and nearly complete \citep{Jester05}.
We use the $z<0.5$ subsample of 87 quasars \citep{BG92}, which was extensively studied 
over a wide range of wavelengths, and is the prime sample of bright AGN. 
The extensive data base allows to explore relations of the radio emission with
a wide range of emission properties.

Of the 87 objects in the \cite{BG92} sample, 71 are defined as radio quiet, based on the \cite{Ke89} 
definition of having $R<10$, where $R\equiv f({\rm 4400\AA})/f({\rm 6~cm})$. 
Of these 71 objects, 62 were detected by \cite{Ke89}
at 5~GHz with the D configuration, where the resolution is 14 arcsec, which generally 
encompasses the entire 
host emission. Of these 62 objects, 53 were in addition detected 
at 5~GHz with the high resolution A configuration of 0.4 arcsec.
We restrict the sample to the 32 objects with detections at a level of 
$>5\sigma$, i.e. a flux $>0.3$~mJy, as some of the suggested $3\sigma$ detections in \cite{Ke89} were found to be non detections in the reanalysis of 
of the same data set by \cite{Miller93}. The $>5\sigma$ detection criterion also 
reduce the error in the derived $5-8.4$~GHz spectral slope. 

The next step is to search for published 8.4~GHz observations of these 32 objects. 
Only 15 of the 32 have published detections with the A configuration \citep{Kukula95, Kukula98, Runnoe18}. 
Three additional objects from the sample are part of the 8.4~GHz variability study of
\cite{Barvainis05}, using mixed configurations. One additional object was observed in the 
C/D configurations by \cite{BL97}, and five more objects by \cite{Barvainis96} in the C configuration.
Observations of two additional objects are found in the {\em VLA} archives, with the A and B configurations, leading to a total of 26 objects with 8.4~GHz observations at al configurations.
Of these 26, 15 were observed in the A configuration, four in a mixed configuration including
A or B, and seven in the C and D configurations. The B, C and D configurations at 8.4~GHz
correspond to angular resolutions of 0.7, 2.5, and 8.6 arcsec,  compared to 0.4 arcsec
in the A configuration. 
The derived slope in these objects is therefore
subject to a possible bias due to a mismatched aperture effect, as the 5~GHz
fluxes of all objects are from the A configuration.
To minimise the aperture effect, we checked the value of the
core dominance parameter $f_c$, defined as the 5~GHz peak flux density in the A configuration to the total 5~GHz flux density
in the D configuration, in the seven objects. We excluded PG~1416$-129$, where $f_c=0.22$ \citep{Ke89}, 
indicating a dominant extended flux, which implies that the available flux in the 8.4~GHz D 
configuration likely includes significant extended flux, which will bias
the derived slope.
In the other 6 objects $1.43\ge f_c\ge 0.77$, which indicates a largely compact 5~GHz source,
and therefore very likely a compact 8.4~GHz source as well. The available C and D configurations 
flux in these objects likely does not include significant extended flux, and should provide a
good estimate of the core flux. The exclusion of PG~1416$-129$ leaves a total of 25 objects.

Another possible source of uncertainty in the spectral slope determination, is variability, indicated
in three of our objects where $f_c>1$. Variability necessarily implies a compact source, less than a fraction of a light year across, given the time difference between the  
A and D configuration observations \citep{Ke89}. Such a compact source, at the
typical 5~GHz luminosity of RQQ, is expected to be optically thick (LB08)
and produce a flat spectrum. Indeed, all three objects show a flat slope. 
Fortunately, the non simultaneity of the 5 and 8.4 GHz observations in these objects did not lead
to an apparent steep spectral slope, which is not possible given their compact sizes.
A compact variable source may just as well induce a wrong $f_c<1$.
We cannot exclude a few objects where a steep slope is a variability artefact. 
To minimise the effect of variability, we use the 5 and 8.4 GHz A configuration 
observations as close in time as possible. In a few objects where repeated observations 
indicated significantly (say a factor of $>2$) variability, we preferred the closer in time 
observations in both bands, if available, even from lower resolution configurations.

An alternative definition of radio loudness was suggested by \cite{Terashima03} based on the 
radio to X-ray flux ratio
$R_X\equiv f({\rm 1~keV})/f({\rm 6~cm})$, where RLQ are defined by having 
$\log R_X>-3.5$\footnote{This value is derived by \cite{Terashima03} for the PG quasars,
compared to $\log R_X>-4.5$ derived for low luminosity AGN.}. 
This criterion is
consistent with our earlier finding (LB08) that   
the RQ PG quasars have $\log R_X\simeq -5$ with a rather small spread. Three of our RQ PGs, PG~1351+640, PG~1700+518,
and PG~2112+059, are RL based on the $\log R_X>-3.5$ criterion. However, all three also show
significant C~IV absorption (PG~1700+518 is a BALQ - broad absorption line quasar),
and have a steeper $\alpha_{ox}$ than found in unabsorbed quasars 
(see Fig.4 in \citealt{LB02}). It is therefore possible that both the X-rays and the UV 
are affected by absorption, which will bias their $R_X$ values to be higher than intrinsic.
We therefore do not remove these three objects from the RQ sample, despite their
having $R_X>-3.5$ values.

Table 1 lists the total sample of the 25 RQ PGs where the 5 and 8.4 GHz spectral slope, 
$\alpha_R$, is derived. The table also
lists various AGN parameters, including the bolometric luminosity, radio luminosity,
$R_X$, the H$\beta$ FWHM, $M_{\rm BH}$, $L/L_{\rm Edd}$, the 
measured 5 and 8.4 GHz fluxes, and $\alpha_R$.
Also listed is the published {\em ROSAT} soft X-ray ($0.2-2$~keV) slope, $\alpha_X$, 
available for 21/25 of the objects. The various relevant references are detailed in the table 
notes. Some of the parameters in the table are used below to study their possible correlations with $\alpha_R$.

As noted above, in this study we explore only the core emission properties.
Eight of the 25 of the objects in the sample have $f_c< 0.5$ \citep{Ke89},
indicating significant flux on scales larger than 0.4 arcsec, which we do not measure.
However, only in two of the eight objects $f_c< 0.3$. Overall, the core radio emission 
dominates ($f_c> 0.5$) in 17/25 of our objects, is comparable ($f_c\sim 0.5$) to the extended 
emission in 6/25, and is significantly smaller ($f_c< 0.5$) only in 2/25.

As a comparison sample, we use the 16 RL PGs. These are selected based on exactly the same 
selection criteria as the 25 RQ PGs. Their emission properties are derived 
from studies of the PG quasars, irrespective of radio loudness. This again guards against any systematic biases in terms of the derived properties of the RQ versus RL PGs.

Table 2 lists the 16 RL PGs, defined by $R>10$ \citep{Ke89}, and
their various emission properties. 
The interpolated $\alpha_R$ near 5 GHz are provided by \cite{Falcke96}, based on
a compilation of literature values, together with single dish {\em Effelsberg} observations.
These observations generally include the radio emission integrated over the host galaxy 
scale or larger, in contrast with the sub kpc integration used above for the RQ PGs.
The different integration scales are required, as the RLQ emission generally extends out to
large scales. In contrast, in the RQQ sample
only the sub kpc scale emission is included, to minimise host contamination.
An alternative approach of measuring $\alpha_R$ in the RLQ on the same sub kpc scale used for
the RQQ, generally reveals only a flat core component. 

\section{The results}

Figure 1 presents the correlations of $\alpha_R$ with the H$\beta$~FWHM, $L/L_{\rm Edd}$ and 
$M_{\rm BH}$, for the 25 RQ PGs (left hand panels), and for the 16 RL PGs (right hand panels).
Note that the three parameters, H$\beta$~FWHM, $L/L_{\rm Edd}$ and 
$M_{\rm BH}$ are not independent, as 
$M_{\rm BH}\propto L^{0.69}\times ({\rm H}\beta~{\rm FWHM})^2$ \citep{DL11}, and 
$L/L_{\rm Edd}\propto L^{0.31}\times({\rm H}\beta~{\rm FWHM})^{-2}$. So, there are only two independent
parameters, either $L$ and H$\beta$~FWHM, or equivalently 
their linear combination in log space, $M_{\rm BH}$ and $L/L_{\rm Edd}$. 
A highly significant correlation of $\alpha_R$ with the H$\beta$ FWHM, or equivalently with 
 $L/L_{\rm Edd}$, is present in the RQ PGs, where $\alpha_R$ gets steeper with decreasing H$\beta$ FWHM, or with increasing $L/L_{\rm Edd}$. 
The Spearman rank order correlation 
coefficient is $r_{\rm S}\simeq 0.67-0.7$ for both correlations, with a probability 
of being random of $p\simeq (1-3)\times 10^{-4}$ (using the two tailed t-test). 
The strong  correlation of $L/L_{\rm Edd}$ with $\alpha_R$  does not necessarily result from
the similarly strong correlation of the H$\beta$ FWHM with $\alpha_R$, together the tight
dependence of $L/L_{\rm Edd}$ on the $\beta$ FWHM. This is evident from the significantly
weaker correlation of $M_{\rm BH}$ with $\alpha_R$ (lower left panel), 
where only a marginal trend exist ($p=0.05$), 
despite the same strong dependence of $M_{\rm BH}$ on the H$\beta$ FWHM.

In contrast with the above correlations for the RQ PGs, the 
RL PGs show different relations. There is only a weak 
trend of increasing $\alpha_R$ with decreasing 
H$\beta$ FWHM, and with increasing $L/L_{\rm Edd}$ (significance $p=0.08-0.11$), 
which is the opposite of 
the relations found in the RQ PGs.
However, a significant correlation is present between $\alpha_R$ and $M_{\rm BH}$
($r_{\rm S}=-0.6856$, $p=0.0034$, i.e $>3\sigma$), of decreasing 
$\alpha_R$ with increasing $M_{\rm BH}$, in contrast with the 
marginal and opposite trend followed by the RQ PGs. 

Figure 2 presents correlations of some additional emission properties of the RQ PGs 
with $\alpha_R$. Of all the optical emission line parameters 
published by \cite{BG92}, the strongest correlation is found 
between $\alpha_R$ and the Fe~II/H$\beta$ line ratio (upper left panel,
$r_{\rm S}=-0.620, p=0.001$). A flat $\alpha_R(>-0.5)$ is always associated with a low Fe~II/H$\beta(<0.4)$, while a steep $\alpha_R(<-0.5)$ is associated with a wide range 
of Fe~II/H$\beta$ values. Similarly, a high Fe~II/H$\beta(>0.6)$
is found only in steep $\alpha_R(>-0.5)$ objects. 
A marginally significant
correlation is present between $\alpha_R$ and the H$\beta$ emission line asymmetry 
(lower left panel, $r_{\rm S}=-0.476, p=0.0162$), where the steep $\alpha_R$ objects tend to have positive asymmetry, i.e. excess blue wing flux \citep{BG92}.

Although the Fe~II/H$\beta$ correlation with $\alpha_R$ is significant, 
it is not independent of the above $\alpha_R$ versus $L/L_{\rm Edd}$
correlation, as Fe~II/H$\beta$ and $L/L_{\rm Edd}$ are  
strongly correlated ($r_{\rm S}=0.777$). Indeed, a partial correlation analysis
of Fe~II/H$\beta$ with $\alpha_R$, at a fixed $L/L_{\rm Edd}$, is insignificant
($r_{\rm S}=0.176$). Similarly, the weaker correlation of the H$\beta$ asymmetry 
with $\alpha_R$ ($r_{\rm S}=-0.476$), becomes insignificant at a fixed $L/L_{\rm Edd}$. 
This suggests there is a single physical mechanism which underlies the relation of
$\alpha_R$ with $L/L_{\rm Edd}$ and the other 
quantities, Fe~II/H$\beta$ and the H$\beta$ asymmetry, which are tightly related
with $L/L_{\rm Edd}$.

Figure 2, upper right panel, explores the relation of $\alpha_R$ and $\alpha_X$ -
the soft X-ray spectral slope, as measured by {\em ROSAT}
at 0.2-2~keV. The value of $\alpha_X$ is available for 21 of the objects 
(see compilation of references in the notes to Table 1).  
The $\alpha_R$ versus $\alpha_X$ 
relation is interesting to explore as $\alpha_X$ is also known to be correlated with
$L/L_{\rm Edd}$ \citep{Laor97, Grupe04, Porquet04, Shemmer06}. A marginally
significant correlation is present between $\alpha_R$ and $\alpha_X$ 
($r_{\rm S}=0.418$, p=0.0589), where the flat
$\alpha_R$ objects have a flat $\alpha_X$ ($>-1.65$). A steep 
$\alpha_R$ is, however, associated with the full range of observed $\alpha_X$ values. 
A marginally significant correlation is also present between $\alpha_X$ and $L/L_{\rm Edd}$  
($r_{\rm S}=-0.443$), which is not as strong as the 
correlation of $\alpha_R$ with $L/L_{\rm Edd}$ ($r_{\rm S}=-0.695$) found here.

Figure 2, lower right panel, shows there is no statistically significant 
relation between $\alpha_R$ and the core dominance parameter, $f_c$ ($r_{\rm S}=-0.077$).
In contrast, the 16 RL PGs show a very strong correlation
($r_{\rm S}=0.726$). Specifically, 4/5 of the flat 
spectra RL PGs have $f_c>0.7$, while all 11 steep spectra objects 
have $f_c<0.4$. This relation implies that an unresolved source, 
i.e. a source smaller than a fraction of a kpc, is generally an optically thick synchrotron source,
and the emission actually comes from a region smaller than a few pc. 
An extended source is always optically thin (steep slope), as expected. 
The tight $\alpha_R$ versus $f_c$ relation 
generally holds in radio selected RLQ, and is interpreted
as a combination of beaming and inclination effects, of otherwise similar
objects. However, in contrast with radio selected AGN, the RL PGs
are selected independently of their radio properties, and it is not clear
that their observed distribution of radio properties can be explained by 
only beaming and inclination, rather than a distribution of intrinsic properties. 

Note that despite the lack of an $\alpha_R$ versus $f_c$ correlation in the RQ PGs, 
the four objects with $f_c>1.1$, PG~1011$-$040, 
PG~1216+069, PG~1351+640, and PG~2304+042, i.e. the objects which are inevitably 
variable, and therefore necessarily compact, do tend to be on the flat $\alpha_R$
side.

There is an equal number of RQ PGs with steep and flat core emission of
the eight objects at $f_c<0.5$. This may well be a true, rather than variability 
effect on $f_c$, as the core
can still be extended only to a fraction of kpc, and be dominated by optically thin
synchrotron source.  The main difference between RLQ and RQQ is that in RLQ the
core emission is generally flat\footnote{excluding the small sub-class of compact 
steep spectrum objects, see \cite{Odea98}.}, while in the RQ PGs, the core 
can be either steep or flat.

\subsection{VLBI observations}

Important input concerning the nature of the radio emission is provided by high resolution 
{\em VLBI} observations of the
mas scale emission. Although extensive {\em VLBI} studies are available for nearby AGN
(e.g. \citealt{Middelberg04} and references therein), most of the objects observed either have
very low accretion rates ($\log L/L_{\rm Edd}<-2$), are obscured AGN, or are RL AGN.
Only a handful of {\em VLBI} studies 
include RQQ, and only a couple of these studies 
\citep{Ulvestad05a, Panessa13} provide observations in more than one band, required
to measure the spectral slope, $\alpha_{\rm mas}$, of the mas scale emission. 
\cite{Ulvestad05a} provide
multifrequency observations for four RQQ. Two belongs to our 
sample, PG~1216+069 which has $\alpha_{\rm mas}=-0.13$, and PG~1351+640
with $\alpha_{\rm mas}=-0.96$, measured between 2.27 and 5~GHz.
One of the two other sources is a bright Seyfert galaxy (Mrk~817), with 
$\alpha_{\rm mas}=-0.60$ between 1.7 and 5 GHz. This AGN is actually 
listed in the PG catalog (PG~1434+590, \citealt{Green86}) but is excluded from the 
final PG quasar sample based on the lack of a quasi-stellar morphology (being host 
dominated). Based on Table 1 in \cite{Vestergaard06} it has $\log L/L_{\rm Edd}=-0.83$,
which is typical in our sample. We therefore include this object in this analysis.
The fourth object is IRAS 07598+6508, which also has a high $L/L_{\rm Edd}$ (see below).
It has $\alpha_{\rm mas}=-1.23$ at 1.67 - 5~GHz.
This is a low ionisation BALQ 
\citep{BM92}, which is likely highly absorbed. The {\em VLA} C/D configuration
gives $\alpha_R=-1.41$  \citep{BL97}. We estimate its $L/L_{\rm Edd}$
based on the H$\alpha$ FWHM (2550~km~s$^{-1}$),
and 5000~\AA\ luminosity \citep{BM92}, and the standard
expressions \citep{Vestergaard06}, which together give $\log M_{\rm BH}=8.18$ and  
$\log L/L_{\rm Edd}=-0.28$, again typical in our sample.

\cite{Panessa13} provide $\alpha_{\rm mas}$ for 23 nearby Seyfert galaxies, most of which
are of extremely low $L/L_{\rm Edd}$, or are type 2 AGN. The only type 1 AGN in their study with 
$\log L/L_{\rm Edd}>-2$ is NGC~4151, where $\log L/L_{\rm Edd}=-1.64$ \citep{Vestergaard06},
and $\alpha_{\rm mas}=0.25$ at 1.7 - 5~GHz. 

We also find two non simultaneous {\em VLBI} observations at different bands for PG~1700+518.
\cite{Blundell98} find a flux of 0.8~mJy with the {\em VLBA} at 8.4~GHz, and \cite{Yang12}
measure a flux of 1.9~mJy at 1.6~GHz using the European {\em VLBI} Network, which together 
imply $\alpha_{\rm mas}=-0.52$ at 1.6 - 8.4~GHz.

Figure 3 compares the position of the above 6 objects in the $\alpha_{\rm mas}$
versus $L/L_{\rm Edd}$ plane, compared to the distribution of $\alpha_R$
versus $L/L_{\rm Edd}$ derived earlier. A dotted line
connects $\alpha_{\rm mas}$ and $\alpha_R$ in the three objects which
are also part of our {\em VLA} sample. Although the {\em VLBI} sample is small,
the two distributions appear to generally overlap. The one somewhat deviant object
is PG~1351+640, with a rather
steep slope of $\alpha_{\rm mas}=-0.96$, despite its low  $L/L_{\rm Edd}$.
This is a rather highly variable object, which varies by a factor of 4 within
3 years at 5 and 15~GHz (see Fig.3 and table 7 in \citealt{BA89}). It is a marginally
RLQ, with $R=4.32$ \citep{Ke89}, and in fact reaches $R>10$ at the high state.
This large amplitude variability is generally not seen in RQQ \citep{Barvainis05}.
So PG~1351+640 appears to be an intermediate object between RQQ and RLQ.

The fraction of the 5~GHz {\em VLA} A configuration core flux ($<0.4$ arcsec), 
which remains 
unresolved on the {\em VLBI} mas scale, is in the range of $0.5-1$, 
indicating that most of the core flux is from a compact source on a pc scale,
rather than an extended source on the sub-kpc scale.

\section{Discussion}

The main result of our study of the 5-8.4~GHz radio core ($<0.4$~arcsec) emission   
of 25 RQ PGs, is the correlation of $\alpha_R$
and $L/L_{\rm Edd}$. Specifically, nearly all RQQ in our sample with $L/L_{\rm Edd}<0.3$ show a flat 
slope ($\alpha_R>-0.5$), and all RQQ with $L/L_{\rm Edd}>0.3$, show a steep slope
($\alpha_R<-0.5$). 

\subsection{Earlier Studies}

Earlier hints for the $\alpha_R$ versus $L/L_{\rm Edd}$ relation can be found 
in \cite{Barvainis96}, who noted
that BALQs, which are believed to have a high $L/L_{\rm Edd}$ 
(e.g. \citealt{Boroson02}), generally have a steep radio slope. In a followup
study of the radio slope of BALQs, \cite{BL97} refute this earlier finding, as they
find some flat spectra BALQs. Their discrepant objects are, PG~1416$-$129, 
which \cite{Green97} found was misclassified as BALQ, and UM~275 and Mrk~231, which are both
RLQ (have $R>10$, based on fluxes from NED), and are therefore not expected to follow 
the $\alpha_R$ versus $L/L_{\rm Edd}$ relation. Thus, the results of \cite{BL97} 
support, rather than refute, the earlier suggestion of \cite{Barvainis96} 
that RQ BALQs have a steep radio slope. 

Another hint can be found in a conference proceeding of \cite{Moran00},
who noted that narrow line Seyfert 1 galaxies, which are generally
high $L/L_{\rm Edd}$ AGN, tend to have a steep radio slope. 
In addition, \cite{Hwang18} recently found that extremely red quasars (ERQs) are characterised 
by a steep radio slope. The ERQs are extreme luminosity quasars, with a median
$L_{\rm bol}\simeq 10^{47}$~erg~s$^{-1}$, and are thus very likely high 
$L/L_{\rm Edd}$ quasars, as also indicated by their emission line properties
\citep{Hamann17}. Thus, the $\alpha_R$ versus $L/L_{\rm Edd}$ relation
may hold over a wide range of luminosities, from the nearby Seyferts level 
to the most luminous high redshift quasars.

A potentially related result is that very low luminosity AGN
with $L/L_{\rm Edd}\ll 10^{-2}$, generally 
have a flat radio slope (see \citealt{Ho08} and references therein).
This may imply that the relation found here of a flat $\alpha_R$ 
down to $L/L_{\rm Edd}\simeq 10^{-1.5}$,
may extend to much lower $L/L_{\rm Edd}$. 

\subsection{Implications}

The direct physical implication of the $\alpha_R$ versus $L/L_{\rm Edd}$ relation 
is that the 5-8.4~GHz synchrotron
source in lower $L/L_{\rm Edd}$ AGN is optically thick, and in higher
$L/L_{\rm Edd}$ AGN is optically thin. An optically thick source at typical RQQ
radio luminosities implies a physical
size $<0.1$~pc (e.g. LB08, eq.22 there), which indicates the radio
originates on the accretion disk scale. The relativistic electrons and magnetic fields
which produce the synchrotron radiation, 
may be produced locally by an outer accretion disc coronal activity, by coronal mass
ejections (CME) from the inner accretion disc corona, or may be associated with
the base of a non-relativistic compact jet, launched from the accretion disk. 

In contrast, synchrotron emission which is optically thin at 5-8.4~GHz
indicates a source larger than $\sim 0.1$~pc. The source can be as large as  
a few hundred pc, which would remain unresolved at our angular resolution of a fraction 
of an arcsec, for our typical $z\sim 0.1$ objects.
An extended optically thin source can be produced by expanding blobs 
of magnetised relativistic electrons, produced by CME events from the innermost disc.
The ejected blobs will be optically thick on small scales, and will inevitably become 
optically thin as they move out, expand, and cool. 
The question is whether they will remain powerful
enough to dominate the emission. Such large scale blobs may also be ejected in a jet. 
The jet needs to be significantly 
sub-relativistic, given the observed constraints on the proper motion 
of the emission features on the pc scale \citep{Middelberg04, Ulvestad05b}.

\subsection{Outflow?}

An extended synchrotron source may also be produced by an AGN driven wind
interaction with the ambient ISM, and the associated shock acceleration. 
Current such modelling for the radio emission in RQQ
\citep{Jiang10, Faucher12, Nims15} address a more extended, kpc scale AGN driven 
wind, which interacts with the host ISM. However, given the significant 
modelling uncertainties, similar scaled down models of a sub-kpc 
wind interaction with ambient gas, may also provide a solution consistent with observations. 
Such a nuclear wind may extend out with detectable emission to larger scales. 
Indeed, radio images of objects
with resolved radio emission often shows a morphology on the
host galaxy scale which appears as an extension of the core emission, rather than a morphology
which follows the host morphology \citep{Ke89, Miller93, Gallimore06}. 
Is the presence of resolved host scale radio emission related to the 
presence of unresolved optically thin core emission? Such a relation is expected
if both are driven by a given nuclear AGN powered wind which extends from the sub-kpc to the kpc scale.

The optically thin radio emission is observed almost exclusively at high
$L/L_{\rm Edd}$, do these objects preferentially show host scale radio emission?
Of the 14 RQ PGs with $L/L_{\rm Edd}>0.3$, 13 show steep $\alpha_R<-0.5$ emission, 
of which 9 (=69\%)
have $f_c\leq 0.77$, indicating significant emission outside the core 
region, i.e. on scales larger than a few hundred pc. However, a similar fraction of 7/11 (=64\%) 
of the flat spectrum RQ PGs also have $f_c\leq 0.77$. Thus, the
presence of significant emission on the host galaxy scale is unrelated to
$L/L_{\rm Edd}$ and to the core $\alpha_R$. A lack of a relation is also suggest by the absence of a significant
correlation between $f_c$ and $\alpha_R$ (Fig. 2, lower right panel).  Steep
slope ($\alpha_R<-0.5$) core emission is found at all values of $f_c$.
Similarly, the lowest values of $f_c\sim 0.3$ is found at all values of $\alpha_R$.
Thus, if the core steep $\alpha_R$ emission is due to a wind/jet, then this outflow 
is not related to the extended host scale emission.  

\subsection{Star Formation?}
Another source for steep core emission is sub-kpc star formation activity. 
Indeed, a high $L/L_{\rm Edd}$ may be associated with a higher specific
star formation rate (SFR) \citep{Sani10, Heckman14}.  
However, can the observed level of the radio luminosity from the core be produced 
by plausible levels SFR? The observed radio luminosity
versus SFR relation (e.g. \citealt{Brown17, Taba17}) and the typical
$L_{\rm R}\sim 10^{39}-10^{40}$erg~s$^{-1}$ in our sample (Table 1, extended with
$\alpha=-1$ to 1.4~GHz) implies a star formation rate of 10-100~M$_{\odot}~{\rm yr}^{-1}$ 
(\citealt{Brown17}, Fig.12 there), or 40-400~M$_\odot~{\rm yr}^{-1}$ 
(\citealt{Taba17}, eq.16 there). This SFR should be confined typically to
 $<0.3$~kpc from the centre (to remain unresolved with the {\em VLA}), which implies a rather extreme 
SFR$>100-1000~M_{\odot}~{\rm kpc}^{-2}~{\rm yr}^{-1}$ at the core.
The Kennicutt-Schmidt law implies an associated surface gas density (e.g. \citealt{KE12}, Fig.11 there)
of $>10^4-10^5~M_{\odot}~{\rm pc}^{-2}$, or a total associated gas mass 
$M_{\rm gas}>3\times 10^{10}-3\times 10^{11}~{\rm M}_\odot$ at the core. 
Can such a high $M_{\rm gas}$ reside at the core?
The implied associated Keplerian velocity on this scale
is $V\sim \sqrt{GM/R}>700-2000~{\rm km}~{\rm s}^{-1}$, which can generally 
be ruled out by the narrow
lines emitted on this scale. The expected continuum 
emission associated with the SF at other bands, and the various 
narrow emission lines luminosities associated with the above SFR \citep{Taba17, Brown17},
may also allow exclusion of such extremely high SFR  
at the core of RQQ (e.g. \citealt{Zakamska14}).

\subsection{Are RQQ a scaled down version of RLQ jets?}

Possibly the simplest interpretation of RQQ is that their radio emission is 
also powered by a relativistic jet, as in RLQ, but their jet power is scaled down by a factor
of $\sim 1000$. In RLQ $\alpha_R$ depends on orientation due to relativistic beaming, 
where edge-on RLQ, with a jet side-view, are steep spectrum lobe dominated, 
and face-on RLQ, with a head-on view of the jet, 
are generally flat spectrum core dominated, e.g. \cite{WB86}. Could $\alpha_R$ in RQQ also be set by inclination?
If true, then face on flat $\alpha_R$ RQQ are expected to have a narrower 
H$\beta$~FWHM, due to a face-on view of the broad line region as well, 
and the steep $\alpha_R$ objects will tend to show a broader
H$\beta$~FWHM due to the edge on view, as observed in RLQ \citep{WB86}. However,
as shown in Fig.1, the RQ PGs show the opposite correlation, where 
flat $\alpha_R$ RQQ, presumably face-on, show broader rather than narrower lines. This
excludes jet collimation together with orientations as the origin of the $\alpha_R$ correlations
in the RQ PGs. 

Although orientation is ruled out, a scaled down non-relativistic jet which emits
isotropically, is not excluded. However, why does $\alpha_R$ from a weak 
non-relativistic jet shows the correlations found here is not addressed by this
scenario. 
 
Since RLQ are generally derived from radio selected samples, while RQQ from optically
selected samples, the observed differences may just reflect the different sample
selection criteria, of otherwise similar objects. To overcome this bias, one needs
to compare RLQ and RQQ with the same selection criteria. 
The RL and RQ PGs selection criteria are the same, and therefore 
the different $\alpha_R$ correlations in the two populations (Fig.1), 
provides strong evidence for significantly different radio emission mechanisms.
The RL PGs show a significant correlation of $\alpha_R$ with $M_{\rm BH}$, while 
the RQ PGs show a marginal and opposite trend. The high $M_{\rm BH}$ RL PGs generally
show a steep slope spatially extended lobe emission.
If a similar scaled down mechanism existed in the RQ PGs, they would have shown the same trend
of $\alpha_R$ with $M_{\rm BH}$ as the RL PGs, even if the lobes were compact and 
spatially unresolved.   

Similarly, the correlation of $\alpha_R$  with $L/L_{\rm Edd}$ in the RQ PGs
is absent in the RL PGs (with a hint of a reverse trend). The RQ PGs at low 
$L/L_{\rm Edd}$  do not produce an extended optically thin radio source. 
In contrast, in RLQ, jet launching and extended radio emission is prevalent at
low $L/L_{\rm Edd}$. Again pointing at a different radio production mechanisms in
RL and RQ quasars.

\subsection{How compact is the optically thin source?}

A direct measurement of the size of the unresolved core component is provided 
by comparing the mas scale flux measured with the {\em VLBI}, with the arcsec flux measured
with the {\em VLA}. In flat spectrum objects the flux generally originates on the mas scale
(see \citealt{Orienti10, Middelberg04}), as theoretically expected for an optically 
thick source. But, how compact
is the radio source in the steep spectrum objects?  Six of our RQ PGs were observed
by \cite{Blundell98} with the {\em VLBA} using a single band 
at 8.4~GHz, of which five show a steep spectrum with the {\em VLA}.
In one steep slope object (PG~1700+517) $>90$\% of the{\it VLA} flux remains unresolved
on the mas scale, while in the other five objects the mas includes $<50$\% of the flux.

Two of the steep spectra objects, PG~1216+069 and PG~1351+640, were also observed by 
\cite{Ulvestad05a} with the {\em VLBA} at 5~GHz, who find that the
mas includes 70\% and 97\% of the {\em VLA} flux, in contrast with fractions of 
14\% and 43\% found by \cite{Blundell98} at 8.4~GHz. These rather discrepant results reflect
the fact that both objects are variable (both have $f_c>1$). This demonstrates the 
potential difficulty in comparing non simultaneous {\em VLA} and mas observations.
The presence of significant variability already shows that the radio
source is compact in both objects, and is consistent with their high $f_c$ values.
  
We conclude that the steep component may be as compact as the pc scale 
(PG~1216+069, PG~1351+640, PG~1700+517), but potentially a significant fraction 
originates on 
larger scales. Clearly, a larger systematic study of the steep spectra quasars
on both the mas and arcsec scale is required to get better constraints on the 
spatial extent of the steep spectrum source. 

Does the $L/L_{\rm Edd}$ versus 
$\alpha_R$ correlation extends to the mas scale? The literature data 
we compiled (Fig.3) on six type 1 AGN with $\log L/L_{\rm Edd}>-2$, of which
three belong to the PG sample, suggests (with significant scatter) that the slope versus 
$L/L_{\rm Edd}$ relation may hold on the pc scale. Multi-band {\em VLBI}
observations of a large and well defined sample of RQQ is required to 
properly address this physically interesting question.

The high brightness temperature of the mas emission, 
typically well above $10^8$~K \citep{Ulvestad05a}, 
 clearly excludes a thermal free-free origin of the  
compact pc scale radio emission. Similarly, variability, which is observed 
in some objects, implies a compact sub pc scale source. 
The presence of variability appears to be similar in amplitude in both steep and flat
slope objects \citep{Barvainis05}, which suggests that the optically thin sources
are also on sub pc scale, and only slightly larger than the optically thick
sources. However, the variability detection is generally only marginal, and
systematic high quality studies of larger samples are required for more
robust conclusions.

\subsection{What are the additional correlations telling us?}

Additional hints are provided by the correlations of $\alpha_R$ and the two
emission line parameters, the Fe~II/H$\beta$ line ratio and the H$\beta$ asymmetry parameter
(Fig.2 left panels). As noted above (section 3), these correlations are 
not independent of the $\alpha_R$ versus $L/L_{\rm Edd}$ correlation, as both parameters are
part of the EV1 set of correlations \citep{BG92}. High EV1 objects
are characterised by a narrow H$\beta$, and thus a high $L/L_{\rm Edd}$, 
high values of Fe~II/H$\beta$, and blue excess asymmetry in the H$\beta$ line profile.

What are these correlations telling us? 
A physical mechanism which may tie together all these trends is a radiation
pressure driven wind. The strength of such a wind depends on the ratio
of radiation pressure to gravity, i.e. $L/L_{\rm Edd}$. The radiation
is coupled to the gas through the metals opacity, and is thus expected to scale as the
Fe~II/H$\beta$ flux ratio, which is likely a metallicity indicator \citep{Shemmer02}.
The blue excess emission of H$\beta$ may indicate an outflowing
component, which is optically thick and 
obscures the receding wind component.
Finally, this outflowing component may either shock, or advect outwards 
magnetised plasma, which produces the extended optically thin steep $\alpha_R$ emission.

A relation may also exists between $\alpha_R$ and $\alpha_X$ (Fig.2, upper right panel).
The flat $\alpha_R$ objects all have $\alpha_X>-1.65$, while the steep
$\alpha_R$ objects show a spread of $\alpha_X$ values.
We note that the X-ray spectral quality for some of the objects is rather low,
and some of the $\alpha_X$ values may be significantly uncertain, or be affected by absorption. 

Why does a compact optically thick radio source tend to be associated with a flat
$\alpha_X$? One may raise the following scenario. 
If the X-ray power-law is produced by an optically thin thermal comptonizing
hot corona (as commonly assumed), then a flatter slope is produced by a 
hotter corona with
a larger optical depth (\citealt{RL79}, eq.7.45b there). If the corona is
heated by magnetic reconnection events, and is depleted by CME, wind, or jet, 
then the lack of an outflow may lead to a 
hotter and larger optical depth corona, which produces a flatter $\alpha_X$.
Such a corona also produces a flatter $\alpha_R$, as the larger column magnetised plasma is more optically thick. 
A CME/wind/jet powered by the corona, may lead to a cooler and lower-column
corona, and thus a steeper $\alpha_X$, while the ejected magnetised plasma
will produce an optically thin steeper $\alpha_R$.

\subsection{Future followups}

This study is based on an archival study of non-simultaneous observations of
25 of the 71 RQ PGs. Clearly, a systematic study of $\alpha_R$, based on simultaneous 
observations of a complete and well defined sample, is warranted. 
The PG sample is a natural place to start. This will allow the exploration of how robust
are the empirical relations found here, and whether additional significant 
relations exist between $\alpha_R$ and the observed wide range of quasar 
emission properties.
In particular, is $\alpha_R$ part of the EV1 set of properties, as suggested in
this study? 

Does the $\alpha_R$ versus $L/L_{\rm Edd}$ relation extend to extremely
low $L/L_{\rm Edd}$ nearby AGN? does it hold in luminous high $z$ AGN? 

Do all RQQ harbour a compact flat spectrum component, as expected if magnetised plasma
is always associated with coronal activity? 
If yes, then the flat component may become dominant at a high enough frequency
also in the steep $\alpha_R$ quasars. To answer this question we started a
{\em VLA} survey of the complete PG quasar sample at 45~GHz. Initial results
(Baldi et al. in preparation) suggest a positive answer, as also suggested by
 earlier high frequency studies of AGN \citep{Behar15, Doi16, Behar18}. If such a component
indeed originates in the accretion disk corona, then at high enough frequency 
($>100$~GHz) the mm emission may originate from small enough scales to overlap 
with the X-ray emitting corona. Correlated mm and X-ray variability may then be 
present \citep{Baldi15}, which will allow to search for the Nuepert effect,
a hallmark of magnetic coronal heating in stellar coronae 
\citep{Neupert68, Guedel02}.

How tightly related are the radio and X-ray emission properties? The radio luminosity versus
the X-ray luminosity relation 
suggests a common coronal origin (LB08). Is there also a corresponding relation 
between $\alpha_R$ and $\alpha_X$, as suggested above? An ongoing {\em XMM-Newton} survey of the PG quasar sample 
(Peretz et al. in preparation) will allow us to expand the $\alpha_R$ versus $\alpha_X$ 
relation to the $0.3-10$~keV range. 
Such a study will allow addressing various questions. For example,
do X-ray absorbed objects also show evidence for radio absorption? 
Since the two absorption mechanisms are quite different (bound-free versus free-free), 
such a study can
provide new information on the absorber properties \citep{Levinson95, Wilson98, Pedlar98}.

\section{Conclusions}

An archival study of the 5-8.4 GHz spectral slope, $\alpha_R$, of the core
radio emission of 25 RQ PGs, reveals a highly significant correlation of $\alpha_R$ 
with $L/L_{\rm Edd}$. All quasars with $L/L_{\rm Edd}>0.3$ have $\alpha_R<-0.5$,
and thus their emission is dominated by an extended optically thin emission,
on pc scale or larger. Most quasars at $L/L_{\rm Edd}<0.3$ have $\alpha_R>-0.5$
indicative of a compact, sub pc scale synchrotron source. 

The RL PGs show a significant correlation of $\alpha_R$ with $M_{\rm BH}$.
Most of the PG RLQ with  $M_{\rm BH}>10^9M_\odot$ have $\alpha_R<-0.5$ and their
radio is dominated by the extended lobe emission, and most RLQ
with $M_{\rm BH}<10^9M_\odot$ have $\alpha_R>-0.5$, and their
radio emission is unresolved.

The different relation of $\alpha_R$ with $M_{\rm BH}$ and $L/L_{\rm Edd}$ 
in the RL and RQ PG quasars indicates that RQQ are not just scaled down RLQ, 
but are rather powered by different mechanisms. 

A possible scenario in RQQ is that the extended optically thin emission,
which dominates at high $L/L_{\rm Edd}$, is associated with an outflow
which becomes significant at high $L/L_{\rm Edd}$, such as
radiation pressure driven wind. At lower $L/L_{\rm Edd}$ this extended component
does not dominate the emission, and only a compact optically thick component,
possibly associated with the coronal emission, is present. 

Another interpretation is suggested by the observed $\alpha_R$ versus 
$\alpha_X$ relation, which may imply that the driving mechanism is CME,
or a magnetic wind, which become more significant at higher $L/L_{\rm Edd}$,
deplete the corona and produce a more extended magnetised outflow.

This study indicates that the radio emission in RQQ is part of the EV1 set of correlations,
and is tightly related to the
emission properties at other bands. Given the drastically improving sensitivity 
of the radio telescopes, from the meter to the sub mm bands, and their vastly higher
angular resolution compared to telescopes in other bands, the radio provides a powerful new tool 
to image a range of physical processes in the innermost parts of RQ AGN. 

\section*{Acknowledgements}

This research was supported by the Israel Science Foundation (grant no. 1561/13).
We thank the referee for useful comments on the paper.

\begin{figure*}
\includegraphics[width=15cm, angle=0]{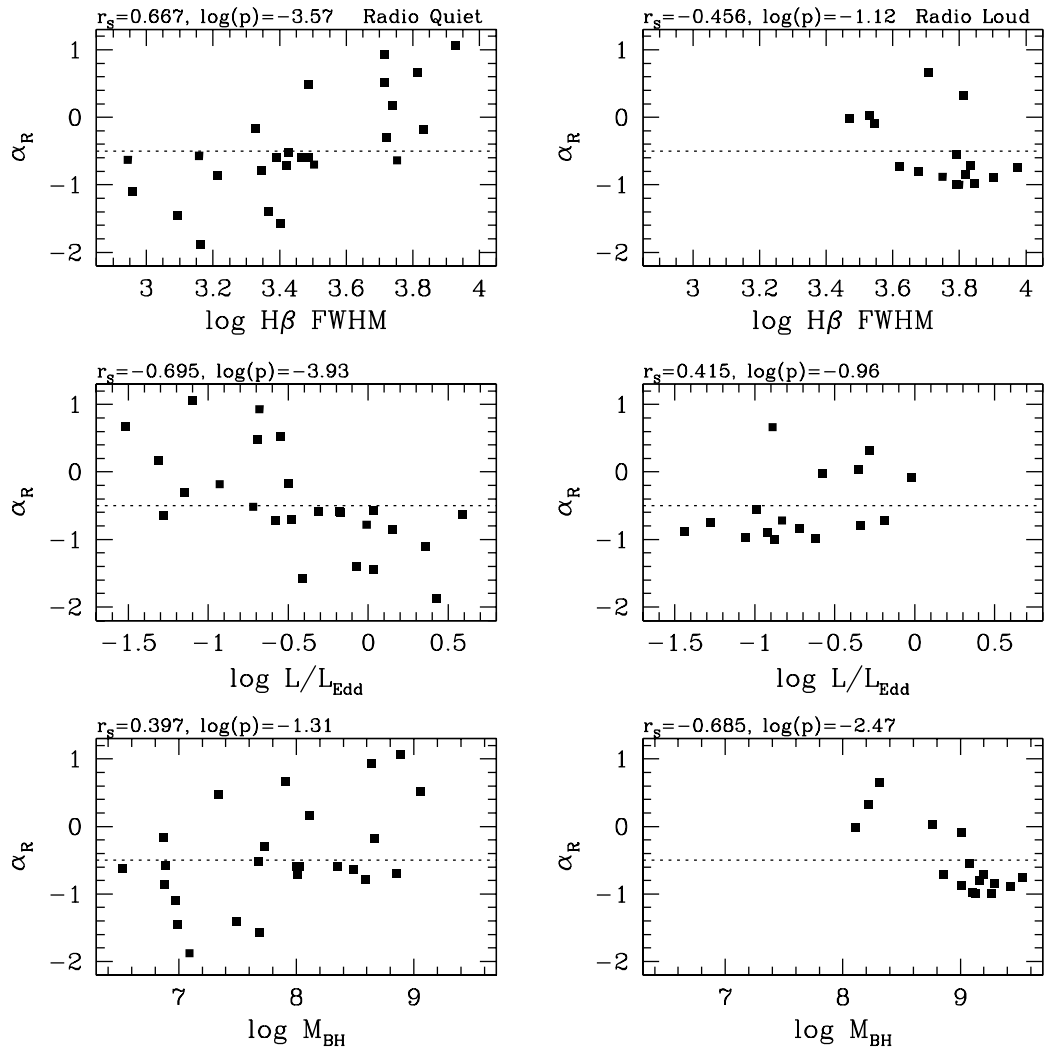}
\caption{The correlation of $\alpha_R$ with the H$\beta$ FWHM, $L/L_{\rm Edd}$, and $M_{\rm BH}$ 
for the RQ PG quasars (left column), and RL PG quasars (right column). The horizontal dotted 
line at $\alpha_R=-0.5$ separates optically thick emission (above) from optically thin emission
(below). The rank order
correlation coefficient ($r_{\rm S}$), and the two tailed significance based on the t-distribution
(p), are indicated above each panel. Note the highly significant p$=(1-3)\times 10^{-4}$ correlation of
$\alpha_R$ with the H$\beta$ FWHM and with $L/L_{\rm Edd}$, and the marginal correlation 
of $\alpha_R$ with $M_{\rm BH}$ (p$\simeq 0.05$) in RQQ. In contrast, in RLQ 
$\alpha_R$ is only marginally correlated, and with an opposite trend, with the H$\beta$ FWHM and with
$L/L_{\rm Edd}$ (p$\simeq 0.1$), and is
significantly correlated with $M_{\rm BH}$ (p$\simeq 0.004$), again with an 
opposite trends. This suggests
different radio production mechanisms in RLQ and RQQ, jets versus potentially an $L/L_{\rm Edd}$
dependent wind (see text).  
}
\label{fig:fig1}
\end{figure*}

\newpage

\newpage
\begin{figure*}
\includegraphics[width=12cm,angle=270]{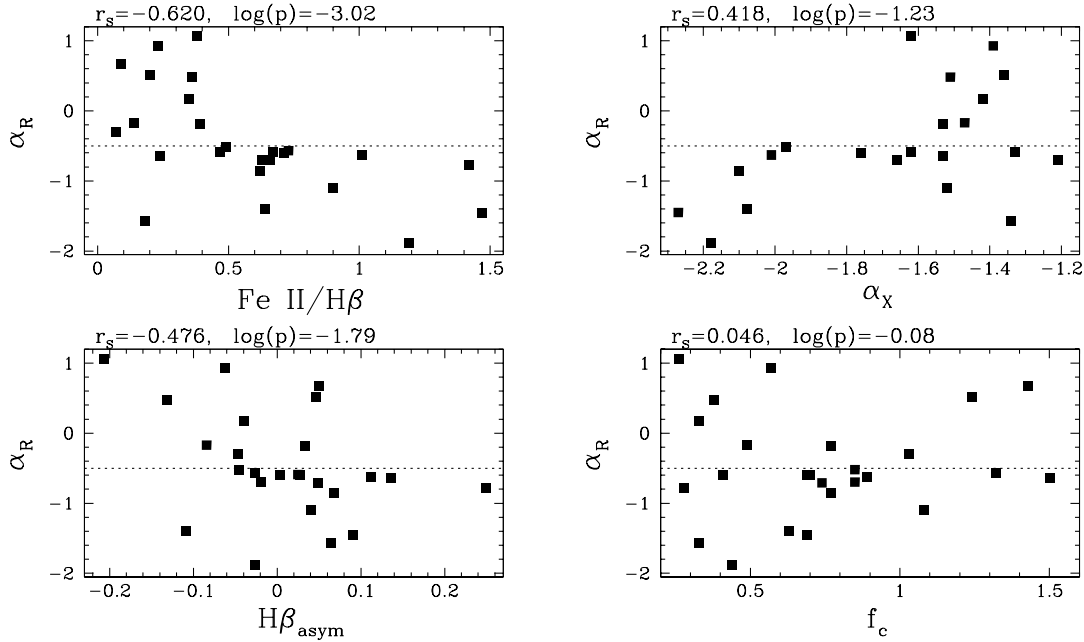}
\caption{Correlations of $\alpha_R$ in the RQ PGs with the FeII/H$\beta$ 
emission line ratio, with the H$\beta$ asymmetry parameter 
(both from \citealt{BG92}), the {\em ROSAT} 0.2-2 keV slope $\alpha_X$, and
the 5 GHz emission core dominance parameter $f_c$. Flat (optically thick) $\alpha_R$ in 
RQQ is associated with a low FeII/H$\beta$, excess red wing emission
in H$\beta$, and a flat $\alpha_X$. These three properties are part of the EV1
set of properties, and are known to correlate with $L/L_{\rm Edd}$.
They may serve as indicators for a radiation pressure driven wind at high $L/L_{\rm Edd}$, 
and an associated extended steep radio emission. 
Note the complete lack of correlation of $\alpha_R$ with $f_{\rm c}$, in contrast with RLQ where 
core dominated ($f_c=1$) sources are generally flat, and extended sources 
are all steep.}
\label{fig:fig2}
\end{figure*}

\newpage

\begin{figure*}
\includegraphics[width=15cm,angle=0]{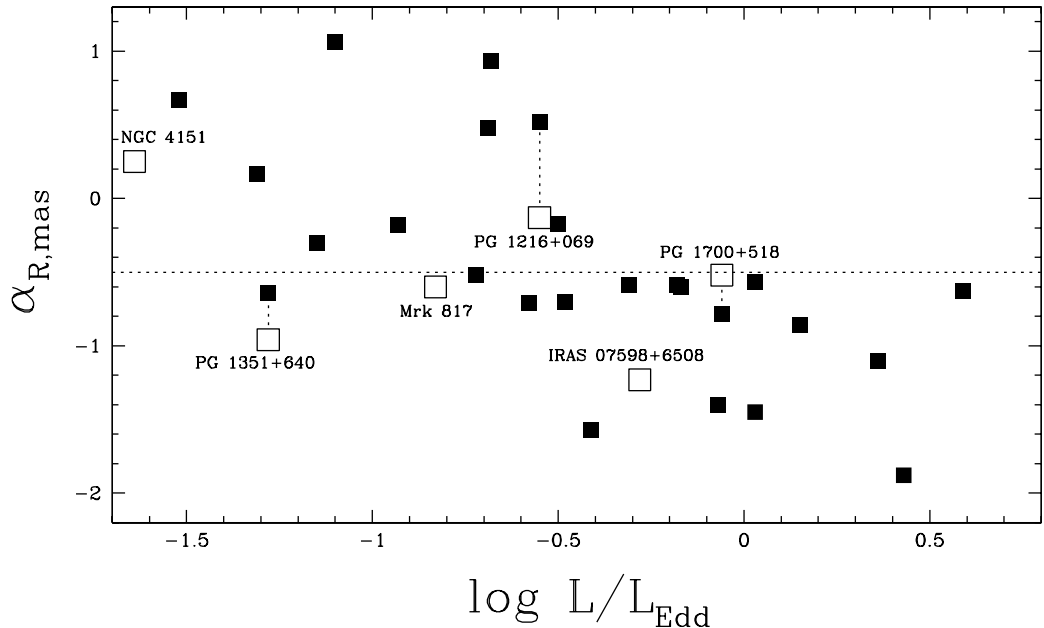}
\caption {{\em VLBI} observations of the mas scale radio slope $\alpha_{\rm mas}$ (open squares)
versus $L/L_{\rm Edd}$, available for three of the 25 RQ PGs, and for three additional type 1 AGN 
with $\log L/L_{\rm Edd}>-2$, on top of the distribution of $\alpha_R$ (filled squares) 
versus $L/L_{\rm Edd}$ (Fig. 1). 
The mas, pc scale, emission appears to follow (with some scatter) the
$\alpha_R$ versus $L/L_{\rm Edd}$ relation, observed at a resolution of a fraction of kpc.}
\label{fig:fig3}
\end{figure*}

\newpage

\begin{table*}
 \centering
 \begin{minipage}{140mm}
  \caption{Radio Quiet PG Quasars}
  \begin{tabular}{@{}rlrlrlrlllrrl@{}}
  \hline
   Name     &  $L_{\rm bol}$ &  L$_{\rm R}$ & $R_X$ & W &  $M_{\rm BH}$ & 
 $\frac{L}{L_{\rm Edd}}$ & $f_5$ & $f_{8.5}$ & Ref. & $\alpha_R$  & $\alpha_X$ & Ref.  \\
(1) &(2) &(3) &(4) &(5) &(6) &(7) &(8) &(9) &(10) &(11) &(12) &(13) \\
 \hline
0003+199   &  45.13 &   38.36 &  $-$5.44 & 1640 & 6.88 & 0.15  & 3.03 & 1.89/A & Ku95 & $-$0.86 & $-$2.10 & G01\\

0049+171   & 44.68$^c$& 38.27 &  $-$5.78 & 5250 & 7.73 & $-$1.15 & 0.66 & 0.56/A/B & Ba05 & $-$0.30 & x &   \\
0050+124   &  45.12 &   38.97 &  $-$5.21 & 1240 & 6.99 &  0.03 & 1.80 & 0.81/A & Ku95 & $-$1.45& $-$2.27&Sc96  \\
0052+251   &  46.06 &   39.35 &  $-$5.58 & 5200 & 8.64 & $-$0.68 & 0.42 & 0.70/A & Ku98 & 0.93 & $-$1.39 & W96\\
0157+001   &  45.93 &   40.13 &  $-$4.10 & 2460 & 8.00 & $-$0.17 & 5.58 & 4.00/A & Ku98 & $-$0.60 & $-$1.76 & UM96\\
0804+761   &  45.82 &   39.88 &  $-$4.90 & 3070 & 8.03 & $-$0.31 & 0.97 & 0.70/A & Ku98 & $-$0.59 & $-$1.33 & WF93 \\
0921+525   &  44.47 &   38.62 &  $-$4.90 & 2120 & 6.87 & $-$0.50 & 1.87 & 1.70/A & Ku98 & $-$0.17 & $-$1.47 & G01 \\
1011$-$040 &  45.02 &   38.09 &  $-$4.51 & 1440 & 6.89 &  0.03 & 0.37 & 0.27/B & Ar & $-$0.57 & x &  \\
1012+008   &  45.53 &   39.50 &  $-$4.56 & 2640 & 8.01 & $-$0.58 & 0.74 & 0.50/A & Ku98 & $-$0.71 & $-$1.66 & W96 \\
1116+215   &  46.27 &   40.20 &  $-$4.63 & 2920 & 8.35 & $-$0.18 & 1.94 & 1.40/A & Ku98 & $-$0.59 & $-$1.62 & G01 \\
1149$-$110 &  44.75 &   38.80 &  $-$5.03 & 3060 & 7.34 & $-$0.69 & 1.00 & 1.30/A & Ku98 & 0.48 & $-$1.51 & W96 \\
1216+069   &  46.61 &   40.62 &  $-$4.48 & 5190 & 9.06 & $-$0.55 & 9.08$^a$ & 12.1/C & Ba96 & 0.52 & $-$1.36 & L97  \\
1351+640   &  45.31 &   40.29 &  $-$3.24 & 5660 & 8.49 & $-$1.28 & 7.02$^a$ & 3.45/C$^b$ & Ba96 & $-$0.64 & $-$1.53 & W96  \\     
1404+226   &  45.21 &   38.67 &  $-$5.10 &  880 & 6.52 &  0.59 & 0.89 & 0.63/C & Ba96 & $-$0.63 & $-$2.01 & UM96 \\
1411+442   &  45.06 &   38.50 &  $-$4.22 & 2670 & 7.68 & $-$0.72 & 0.52 & 0.39/C/D & BL97 & $-$0.52 & $-$1.97 & L97 \\
1426+015   &  45.84 &   39.30 &  $-$5.28 & 6820 & 8.67 & $-$0.93 & 0.84$^a$ & 0.76/C & Ba96 & $-$0.18 & $-$1.53 & WF93 \\
1440+356   &  45.62 &   38.94 &  $-$5.60 & 1450 & 7.09 &  0.43 & 0.73 & 0.26/A & Ar & $-$1.88 & $-$2.18 & G01 \\
1448+273   &  45.43$^c$ &   38.58 &  $-$5.28 &  910 & 6.97$^c$ &  0.36 & 1.34$^a$ & 0.73/C & Ba96 & $-$1.10 & $-$1.52 & G01 \\
1501+106   &  44.90 &   38.92 &  $-$4.72 & 5470 & 6.97 & $-$1.31 & 0.50 & 0.55/A & Ku95 & 0.17 & $-$1.42 & G01  \\
1612+261   &  45.38 &   40.02 &  $-$4.51 & 2520 & 7.69 & $-$0.41 & 1.66 & 0.70/A & Ku98 & $-$1.57 & $-$1.34 & Sc96  \\
1613+658   &  45.89 &   39.55 &  $-$5.34 & 8450 & 8.89 & $-$1.10 & 0.78 & 1.40/A & Ku98 & 1.06 & $-$1.62 & G01\\
1700+518   &  46.63$^a$&40.75 & $-1.74^c$ &2210 & 8.59$^c$ & $-$0.06 & 1.32$^e$ & 0.86/A$^e$ & Ru18 & $-$0.78 & x &   \\
2112+059   &  46.47 &   40.51 &  $-$3.36 & 3190 & 8.85 & $-$0.48 & 0.76 & 0.52/B/D & Ba05 & $-$0.70 & $-$1.21 & W96  \\  
2130+099   &  45.52 &   39.00 &  $-$5.35 & 2330 & 7.49 & $-$0.07 & 1.30 & 0.60/A & Ku98 & $-$1.40 & $-$2.08 & G01 \\
2304+042   &  44.49 &   38.14 &  $-$5.75 & 6500 & 7.91 & $-$1.52 & d & 0.65/A/D & Ba05 & 0.67 & x &  \\

\hline
\end{tabular} \\
Note. -  Col. (1): PG object designation; 
col. (2): Log of bolometric luminosity in erg~s$^{-1}$ \citep{DL11}; 
col. (3) Log of the radio luminosity in observed frame 5~GHz in erg~s$^{-1}$ \citep{Ke89};
col. 4) Log of the radio to X-ray luminosities, taken from LB08 (Table 1 there); 
col. (5) The broad H$\beta$ line FWHM in km~s$^{-1}$ \citep{BG92};
col. (6) Log of the BH mass in units of $M_{\odot}$ \citep{DL11};
col. (7) Log of the luminosity in Eddington units, based on columns (2) and (6);
col. (8) The unresolved (i.e. peak) flux at 4.89~GHz, observed in the A configuration ;
col. (9) The unresolved (i.e. peak) flux at 8.4~GHz, and the configuration used;
col. (10) The reference for the 8.4~GHz observation;
col. (11) The value of $\alpha_R$ based on columns (9) and (10);
col. (12) The {\em ROSAT} $0.2-2$~keV slope, $\alpha_X$, corrected for the foreground Galactic absorption;
col. (13) The reference for the value of $\alpha_X$.

a - The flux at the D configuration in \cite{Barvainis96} and in \cite{Ke89} are significantly different, which indicates variability. 
We therefore measure $\alpha_R$ using the lower resolution but nearly simultaneous  
observations of both bands in \cite{Barvainis96}. 

b - Flux at 14.9~GHz.

c - Estimated as $9\nu L_{\nu}(5100{\rm \AA}$), based on the mean SED of 
\cite{Richards06}, where $\nu L_{\nu}(5100{\rm \AA}$)
is from \cite{Vestergaard06}.

c - Taken from \cite{Vestergaard06}.

d - The \cite{Barvainis05} observations indicate large variability (more than a factor of two in 
less than half a year). We therefore use the slope provided by \cite{Barvainis05}, derived 
from lower resolution, but nearly simultaneous observations.

e - Peak core fluxes (western source) reported in \cite{Runnoe18}.

x - No published {\em ROSAT} detection, or the S/N is too low.

References: 
Ar   - An archive search; 
Ba96 - \cite{Barvainis96}; 
Ba05 - \cite{Barvainis05};
BL97 - \cite{BL97}; 
G01  - \cite{Grupe01}; 
Ku95 - \cite{Kukula95}; 
Ku98 - \cite{Kukula98}; 
L97  - \cite{Laor97}; 
Ru18 - \cite{Runnoe18}; 
Sc96 - \cite{Schartel96}; 
UM96 - \cite{UM96}; 
W96  - \cite{Wang96}; 
WF93 - \cite{Walter93}. 

\end{minipage}

\end{table*}

\begin{table*}
 \centering
 \begin{minipage}{140mm}
  \caption{Radio Loud PG Quasars}
  \begin{tabular}{@{}cllllllrll@{}}
  \hline
   Name     &  $L_{\rm bol}$ &  L$_{\rm R}$ & $R_X$ & W &  $M_{\rm BH}$ & 
 $L/L_{\rm Edd}$ & $\alpha_R$   \\
(1) &(2) &(3) &(4) &(5) &(6) &(7) &(8)   \\
 \hline

0003+158 & 46.92 & 42.99 & $-$2.56 & 4760 &  9.16 & $-$0.34 & $-$0.80  \\    
0007+106 & 45.52 & 41.93 & $-2$.56 & 5100 &  8.31 & $-$0.89 & 0.66 \\
1004+130 & 46.49$^a$ & 42.70 & $-1.78$ & 6300 & 9.27$^b$ & $-$0.88 & $-$1.00   \\
1048$-$090 & 46.57 & 42.97 & $-2.19$ & 5620 &  9.01 & $-$1.44 & $-$0.88  \\    
1100+772 & 46.61 & 42.92 & $-$2.32 & 6160 &  9.13 & $-$0.62 & $-$0.99  \\  
1103$-$006 & 46.19 & 42.88 & $-2.02$ & 6190 &  9.08 & $-$0.99 & $-$0.55  \\    
1226+023 & 47.09 & 43.97 & $-1.70$ & 3520 &  9.01 & $-$0.02 & $-$0.09 \\  
1302$-$102 & 47.04 & 42.99 & $-2.10$ & 3400 &  8.76 & $-$0.35 & 0.04  \\    
1309+355 & 45.63 & 41.05 & $-2.84$ & 2940 &  8.11 & $-$0.58 &  $-$0.02 \\    
1425+267 & 46.35 & 41.82 & $-2.39$ & 9410 &  9.53 & $-$1.28 & $-$0.75  \\   
1512+370 & 47.11 & 42.63 & $-2.40$ & 6810 &  9.20 & $-$0.19 & $-$0.71  \\   
1545+210 & 46.14 & 42.91 & $-2.23$ & 7030 &  9.10 & $-$1.06 & $-$0.98  \\      
1704+608 & 46.67 & 43.29 & $-1.35$ & 6560 &  9.29 & $-$0.72 &  $-$0.84  \\   
2209+184 & 46.02 & 41.45 & $-2.49$ & 6500 &  8.22 & $-$0.28 &  0.32   \\    
2251+113 & 46.13 & 43.06 & $-1.03$ & 4160 &  8.86 & $-$0.83 & $-$0.72  \\     
2308+098 & 46.61 & 42.76 & $-2.59$ & 7970 &  9.43 & $-$0.92 & $-$0.89  \\    

\hline
\end{tabular} \\
Note. -  Cols. (1)-(7): As in Table 1; 
col. (8) From \cite{Falcke96}.

a - Estimated as $9\nu L_{\nu}(5100{\rm \AA}$), based on the mean SED of 
\cite{Richards06}, where $\nu L_{\nu}(5100{\rm \AA}$)
is from \cite{Vestergaard06}.

b - From \cite{Vestergaard06}.

\end{minipage}

\end{table*}

\end{document}